\documentclass[12pt,a4paper]{elsarticle}

\usepackage{amsmath}
\usepackage{amssymb}

\usepackage{graphicx}
\usepackage{mdframed}

\usepackage{algorithm}
\usepackage{algorithmic}
\usepackage{fullpage}

\newtheorem{theorem}{Theorem}

\newtheorem{lemma}{Lemma}

\newproof{proof}{Proof}

\renewcommand{\H}{\mathcal{H}}
\renewcommand{\S}{\mathcal{S}}
\renewcommand{\P}{\mathcal{P}}

\newcommand{\C}{\mathcal{C}}
\newcommand{\N}{\mathbb{N}}

\newcommand{\F}{\mathcal{F}}

\newcommand{\T}{\mathcal{T}}
\newcommand{\M}{\mathcal{M}}

\newcommand{\xset}{X}
\newcommand{\yset}{Y}
\newcommand{\lab}{l}
\newcommand{\ppart}{\mathcal{P}}
\newcommand{\mmodel}{\mathcal{M}}
\newcommand{\chvd}{\textsc{ChVD}}
\newcommand{\dchvd}{\textsc{D-ChVD}}
\newcommand{\dchvdb}{\textsc{D-ChVD2}}
\newcommand{\fvs}{\textsc{FVS}}

\begin{document}

\begin{frontmatter}
\title{Towards a faster parameterized algorithm for Chordal Vertex Deletion}
\author{Sylvain Guillemot}
\ead{guillemo@free.fr}

\begin{abstract}
The \textsc{Chordal Vertex Deletion} (\chvd{}) problem asks whether an input graph $G$ can be made chordal by removing at most $k$ vertices. When $k$ is taken as a parameter, this problem is known to be fixed-parameter tractable \cite{M10,CM16}. Yet, it remains an open question whether \chvd{} admits a single-exponential FPT algorithm, i.e. an algorithm with running time $O(\alpha^k |V(G)|^c)$ for some constant $\alpha$ independent of $k$. Towards this goal, we describe an FPT algorithm for the compression version of \chvd{} with the following specification. Given an instance $I = (G,k,X)$ of the disjoint \chvd{} compression problem such that $|X| = k+1$, $G - X$ is chordal and its longest chordless path has length $L$, we solve $I$ in time $O((L+1)^{3k} k^2 |V(G)|^4)$.
\end{abstract}
\end{frontmatter}

\section{Introduction}

\textbf{Context.} Let $\C$ be a hereditary class of graphs. The \textsc{$\C$ Vertex Deletion} problem asks, given a graph $G$ and an integer $k$, whether there is a set $X \subseteq V(G)$ of size at most $k$ such that $G - X$ belongs to $\C$. When the class $\C$ is polynomial-time recognizable, it is a logical next step to ask whether the \textsc{$\C$ Vertex Deletion} problem is fixed-parameter tractable for parameter $k$. This is always the case when $\C$ is characterized by a finite number of obstructions \cite{C96}, but the question is non-trivial for classes having unbounded obstructions. In this setting, there are many positive results (see e.g. \cite{W17} for some recent results) but also a number of negative results \cite{L08,HHJKV13} as well as prominent classes for which the question remains open.

Perhaps the simplest and most studied problem in this category is the \textsc{Feedback Vertex Set} (\fvs{}) problem, which corresponds to \textsc{$\C$ Vertex Deletion} when $\C$ is the class of forests. The problem is a test bed for several standard approaches in parameterized complexity, such as bounded-search, treewidth and iterative compression. The \fvs{} problem admits several single-exponential FPT algorithms \cite{CNPPRW22,LN22,CFLLV08,CCL015,KP14}. Arguably, the class of chordal graphs can be seen as an extension of forests that enjoy similar characterizations: the existence of elimination orderings \cite{FG65,R70}, the layout of the separators provided by the clique tree \cite{D61,G74}, and the obstructions provided by chordless cycles. In spite of this, the fixed-parameter tractability of \textsc{Chordal Vertex Deletion} (\chvd{}) proved much more challenging than \fvs{}. The first FPT algorithm for the \chvd{} problem, due to \cite{M10}, uses treewidth reduction and Courcelle theorem. An improvement due to \cite{CM16} gives a bounded-search $O^*(k^{O(k)})$ time algorithm \footnote{We use the notation $O^*$ to hide polynomial factors, i.e. $O^*(f(k)) = f(k) n^{O(1)}$.} for a more general problem allowing edge editing in addition to vertex deletion. Polynomial kernels for \chvd{} were also obtained by \cite{JP18,ALMSZ19}. It should be noted that the previous algorithms use iterative compression, which is now a standard technique in FPT algorithms for graph modification problems. Since the \fvs{} problem admits single-exponential FPT algorithms using iterative compression, this raises the question of whether the same holds for the more general \chvd{} problem.\\

\textbf{Our result.} While we do not resolve this question, we make progress towards a positive resolution. We recall that in the setting of iterative compression, we have to solve the \textsc{Disjoint \chvd{} Compression} (\dchvd{}) problem. An instance of \dchvd{} is a tuple $I = (G,k,X)$ consisting of a chordal graph $G$, of an integer $k$, and of a chordal modulator \footnote{A chordal modulator of $G$ ia set $X \subseteq V(G)$ such that $G - X$ is chordal.} $X \subseteq V(G)$ of size $k$. The instance $I$ is positive iff there exists a chordal modulator $S \subseteq V(G)$ disjoint of $X$ and having size at most $k-1$. We let $L$ denote the maximum length of a chordless path in $G - X$. Since a chordless cycle in $G$ can have length at most $(L+1)k$, we can trivially solve \dchvd{} in $O^*(((L+1)k)^k)$ time by branching. However, improving this to single-exponential proves challenging already when $L = 1$, i.e. when $G - X$ is edgeless. In this paper, we present the following algorithmic result.

\begin{theorem} \label{thm1} The \dchvd{} problem can be solved in $O((L+1)^{3k} k^2 |V(G)|^4)$ time.
\end{theorem}

\textbf{High-level description.} Let us explain the high-level idea behind our algorithm. We solve the following auxiliary problem \dchvdb{}. An instance of the problem is a tuple $I = (G,k,X,Y)$ where $X,Y$ are disjoint subsets of $V(G)$ and $|X| = k$. A solution for instance $I$ is a set $S \subseteq V(G)$ disjoint from $X \cup Y$ such that $|S| < k$ and $G - S$ is chordal. Here, $X$ is the original chordal modulator, and $Y$ is a set of \emph{forced} vertices which must be absent from a solution. We also let $W' = X \cup Y$, $Z = V(G) - W'$ and $V' = Y \cup Z$. We solve \dchvdb{} by branching in the following way. If $k = 0$ or $G[W']$ is not chordal, we reject; if $G$ is chordal, we accept. Otherwise, we identify a specific chordless path $P$ in $G-X$ called a \emph{candidate path}, and we branch on at most $L+1$ possibilities: $(B_1)$ for each $v \in V(P) \cap Z$, we take $v$ in a solution, $(B_2)$ we add $V(P)$ to $Y$. There are two types of candidate paths, called paths of \emph{type 1} and \emph{type 2}.

We proceed to define these two types of candidate paths. 
Let us first define the candidate paths of type 1. A \emph{hole} in $G$ is a chordless cycle $H$ of length at least 4. A \emph{$V'$-fragment} of $H$ is a maximal subpath of $H$ included in $V'$. Recall that since $X$ is a chordal modulator, we have $V(H) \cap X \neq \emptyset$ and thus $H$ cannot be included in $V'$. A candidate path $P$ is of \emph{type 1} if there is a hole $H$ in $G$ such that (a) $P$ is a $V'$-fragment of $H$, (b) for each $Q$ $V'$-fragment of $H$ distinct from $P$, we have $V(Q) \subseteq Y$. Let us now define the candidate paths of type 2. Consider a pair $p = (x,x')$ with $x,x' \in X$ non adjacent. We say that $p$ is \emph{$Y$-connected in $I$} iff there is $Q$ chordless $(x,x')$-path in $G[W']$ such that $V(Q) \cap Y \neq \emptyset$. We let $\S_I$ be the set of pairs $(x,x')$ with $x,x' \in X$ non adjacent and that are not $Y$-connected in $I$. A candidate path $P$ is of \emph{type 2} iff $V(P) \subseteq V'$ and there exists $(x,x') \in \S_I$ such that $xPx'$ is a chordless path in $G$.\\

\textbf{Specifics of the algorithm.} The choice of a candidate path at each step is made by the following criterion. Recall that if $G = (V,E)$ is chordal, an enumeration $\sigma = v_1 v_2 \ldots v_n$ of $G$ is a \emph{perfect elimination ordering} of $G$ iff for each $p,q,r \in [n]$ with $p,q > r$, if $v_p v_r \in E$ and $v_q v_r \in E$ then $v_p v_q \in E$. We suppose in the following that we have $\sigma$ fixed perfect elimination ordering of $G[V']$. We then define the \emph{labelling} $l$ by $\lab(v_i) = i$ for each $i \in [n]$, and given a set $S \subseteq V'$, we let $\lab(S) = \max_{v \in S} \lab(v)$. If there is a candidate path of type 1, we choose one arbitrarily. On the other hand, if there is a candidate path of type 2, we choose one such path $P$ with $\lab(V(P))$ minimum. We show in Section \ref{section4} that finding such a candidate path can be carried out in polynomial time. We note that while there can be an exponential number of such paths, we can give a compact description of them that allows to find one in $O(k^2 n^4)$ time.

Let us explain the rationale behind the definition of candidate paths of type 2. Let us suppose that $G[W']$ is chordal, and consider $H$ hole in $G$. In this case, $H$ must contain a $V'$-fragment $P$ such that $V(P) \cap Z \neq \emptyset$. Moreover, (a) either $P$ is a candidate path of type 1, or (b) there exists $x,x' \in X$ nonadjacent such that $xPx'$ is a subpath of $H$. However, in case (b), $P$ may fail to be a candidate path of type 2 because $(x,x') \notin \S_I$. In this case, since $x,x'$ are $Y$-connected in $I$, we find a chordless $(x,x')$-path $Q$ in $G[W']$ containing a vertex of $Y$. We argue that for a proper choice of $H$ and $P$, we may then ``shorten'' the hole $H$ using the path $Q$ to obtain $H'$ hole in $G$ such that $|V(H') \cap Z| < |V(H) \cap Z|$. This is the purpose of Theorem \ref{thm3} proved in Section \ref{section5}. The criterion for choosing a candidate path is essential here, and ultimately it guarantees that as long as $G$ is not chordal we may find a candidate path of type 1 or 2.\\

\textbf{The measure.} Obviously, in each branching step $(B_1)$ the parameter $k$ drops by 1, but it is not immediate whether we make progress in branching step $(B_2)$. Let us explain this in detail. Observe that if $P$ is a candidate path of type 1, the branch $(B_2)$ will directly reject as after adding $V(P)$ to $Y$ we obtain $H$ hole in $G[X \cup Y]$. On the other hand, if $P$ is a candidate path of type 2 and $(B_2)$ considers the new instance $I'$, we have a pair $(x,x') \in \S_I$ that was not $Y$-connected in $I$ but becomes $Y$-connected in $I'$. This allows us to define a measure $s(I)$ which drops by at least 1 in each branch $(B_2)$, so that the global measure $m(I) = s(I) + k$ drops by at least 1 in each branch $(B_1)$ or $(B_2)$. A crude upper bound gives $s(I) = O(k^2)$, but we can actually give a bound $s(I) \leq 2k$ as shown by Theorem \ref{thm4} in Section \ref{section6}. This yields a global measure $m(I) \leq 3k$ that drops by at least 1 in each branch. Since we branch in at most $L+1$ directions at each step, this readily gives the claimed $O^*((L+1)^{3k})$ running time.

\section{\label{section2}Definitions}

We use the notation $[n]$ to denote the set of integers from $1$ to $n$, and the notation $\N$ to denote the set of integers.\\

\textbf{Graphs.} We denote a graph by $G = (V,E)$ where $V$ is its set of vertices and $E$ is its set of edges. We use the notation $V(G)$, resp. $E(G)$, to refer to the set $V$, resp. $E$. We say that $u,v \in V$ are \emph{adjacent} in $G$ iff $uv \in E$. Let $G' = (V',E')$ be another graph. We say that $G'$ is a \emph{subgraph} of $G$ iff $V' \subseteq V$ and $E' \subseteq E$. We say that $G'$ is an \emph{induced subgraph} of $G$ iff $V' \subseteq V$ and $E' = \{ e \in E : e \subseteq V' \}$. Given $S \subseteq V$, we denote by $G[S]$ the induced subgraph of $G$ with vertex set $S$, and we denote by $G-S$ the induced subgraph of $G$ with vertex set $V-S$. Given $S \subseteq V$, the \emph{contraction of $S$} in $G$ produces the subgraph $G / S$ with vertex set $V' = V - S + \{z\}$ and with edge set $E' = \{ uv \in E : u,v \in V-S \} \cup \{ uz : u \in V-S \text{ and there exists } v \in S \text{ such that } uv \in E \}$.

A \emph{walk} in $G$ is a sequence $W = z_1 z_2 \ldots z_L$ of vertices such that $z_i z_{i+1} \in E$ holds for each $i \in [L-1]$; we denote  $V(W) = \{z_1,z_2,\ldots,z_L\}$ the set of vertices in $W$, $I(W) = \{z_2,\ldots,z_{L-1}\}$ the set of internal vertices in $W$, and $E(W) = \{ z_i z_{i+1} : 1 \leq i < L \}$ the set of edges in $W$. If the vertices $z_1,\ldots,z_L$ are distinct, then $W$ is a \emph{path} in $G$ of length $|W| = L$. If the vertices $z_1,\ldots,z_{L-1}$ are distinct and $z_L = z_1$, then $W$ is a \emph{cycle} in $G$ of length $|W| = L-1$. A \emph{chord} of $W$ is an edge of the form $z_i z_j$ with $i,j$ not consecutive; we say that $W$ is \emph{chordless} if it has no chord.  Given two vertices $u,v \in V$, an \emph{$(u,v)$-path} in $G$ is a path in $G$ starting with $u$ and ending with $v$. If $W = z_1 z_2 \ldots z_L$ is a path or a cycle and $1 \leq i \leq j \leq L$, we define the \emph{subpath} $W[z_i,z_j]$ as the path $z_i z_{i+1} \ldots z_j$.

Given $S \subseteq V$, we define the \emph{neighborhood} $N_G(S) = \{ u \in V-S : \text{ there exists an edge }$ $ uv \in E \text{ with } v \in S \}$, and we define the \emph{extended neighborhood} $N_G[S] = S \cup N_G(S)$. If $v \in V$, we abbreviate $N_G(\{v\})$ by $N_G(v)$, and $N_G[\{v\}]$ by $N_G[v]$. A \emph{clique} of $G$ is a set of pairwise adjacent vertices of $G$. A \emph{separation} of $G$ is an ordered partition $(A,S,B)$ of $V$ such that $A,B$ are nonempty and there is no edge $ab \in E$ with $a \in A, b \in B$. The graph $G$ is \emph{disconnected} iff it admits a separation of the form $(A,\emptyset,B)$, and it is \emph{connected} otherwise. A \emph{connected component} of $G$ is a maximal connected subgraph of $G$. A graph is a \emph{tree} iff it is connected and contains no cycle.\\

\textbf{Chordal graphs.} A \emph{hole} in $G$ is a chordless cycle in $G$ of length at least 4. The graph $G$ is \emph{chordal} iff there is no hole in $G$. A vertex $v \in V$ is \emph{simplicial} in $G$ iff $N_G[v]$ is a clique of $G$. Let $\sigma = v_1 v_2 \ldots v_n$ be an ordering of $V$. We say that $\sigma$ is a \emph{perfect elimination ordering} of $G$ iff for each $i \in [n]$, $v_i$ is simplicial in $G[\{v_i,v_{i+1},\ldots,v_n\}]$. Equivalently, $\sigma$ is a perfect elimination ordering iff for each $p,q,r \in [n]$ with $p,q > r$, if $v_p v_r \in E$ and $v_q v_r \in E$ then $v_p v_q \in E$. It is known that a graph is chordal iff it admits a perfect elimination ordering \cite{FG65,R70}. Moreover, finding a perfect elimination ordering of a chordal graph can be done in linear time \cite{RTL76}.

A \emph{tree family} is a tuple $\T = (T, \{T_v : v \in V \})$ where $T$ is a tree and each $T_v$ is a subtree of $T$, namely a connected subgraph of $T$. We say that $\T$ is a \emph{clique tree} of $G$ if the following condition holds: for each $u,v \in V$, we have $uv \in E$ iff $V(T_u) \cap V(T_v) \neq \emptyset$. It is known that a graph is chordal iff it admits a clique tree \cite{R70,G74}. Suppose that $\T$ is a clique tree of $G$. For each $x \in V(T)$, we let $C^{\T}_x = \{ v \in V : x \in V(T_v) \}$, then each set $C^{\T}_x$ is a clique of $G$, and each maximal clique of $G$ has the form $C^{\T}_x$ for some $x \in V(T)$. Consider an edge $e = xy$ of $T$, and let $C$ (resp. $C'$) be the connected component of $T - e$ containing $x$ (resp. $y$). We then define the separation $S^{\T}_{xy} = (A,S,B)$ of $G$ where $A = \{ v \in V : T_v \subseteq C \}, B = \{ v \in V : T_v \subseteq C' \}$ and $S = \{ v \in V : T_v \text{ contains } x,y \}$. 

\section{\label{section3} The algorithm}

\subsection{Auxiliary results}

For the needs of the algorithm, we consider the following auxiliary problem where a set $Y$ of \emph{forced} vertices is provided in addition to the chordal modulator $X$.\\

\fbox{
\parbox{0.9 \textwidth}{
\textbf{Name:} \dchvdb{}\\
\textbf{Input:} an instance $I = (G, k, X, Y)$ where $G = (V,E)$ is a graph, $k \in \N$ is an integer, $X,Y$ are disjoint subsets of $V$ such that $|X| = k$ and $G - X$ is chordal\\
\textbf{Solution:} a set $S \subseteq V$ disjoint from $X \cup Y$ such that $|S| < k$ and $G - S$ is chordal
}}\\

Fix an instance $I = (G, k, X, Y)$ of \dchvdb{}. We let $W' = X \cup Y$, $Z = V - W'$ and $V' = Y \cup Z$. We assume in the following that we have $\sigma = v_1 \ldots v_n$ fixed perfect elimination ordering of $G[V']$. We define the labelling $l$ by $\lab(v_i) = i$ for each $i \in [n]$. Given $S \subseteq V'$ we let $\lab(S) = \max_{v \in S} \lab(v)$. If $P$ is a path we abbreviate $\lab(V(P))$ as $\lab(P)$. If $H$ is a hole in $G$ and $S \subseteq V$, a \emph{$S$-fragment} of $H$ is a maximal subpath of $H$ included in $S$. Given a set $\F$ of $S$-fragments of $H$, a fragment $P \in \F$ is \emph{$l$-minimal} iff $l(P) < l(P')$ for each $P' \in \F - \{P\}$.

Given two vertices $x,x' \in X$ and a set $S \subseteq V$, we say that $x,x'$ are \emph{$S$-connected in $G$} iff there is $Q$ chordless $(x,x')$-path in $G$ containing an internal vertex in $S$. If $y \in V$, we use the formulation \emph{$y$-connected} to mean $\{y\}$-connected. Note that $x,x'$ are $S$-connected in $G$ iff they are $y$-connected in $G$ for some $y \in S$. Let $\P_I$ be the set of pairs $(x,x')$ such that $x,x' \in X$ and $x,x'$ are not adjacent, and let $\S_I$ be the set of pairs $(x,x') \in \P_I$ such that $x,x'$ are not $Y$-connected in $G[W']$.

Let $P$ be a chordless path in $G[V']$. We say that $P$ is a \emph{candidate path of type 1} (in $I$) iff there is $H$ hole in $G$ such that (a) $P$ is a $V'$-fragment of $H$, (b) for each $Q$ $V'$-fragment of $H$ distinct of $P$, we have $V(Q) \subseteq Y$. We say that $P$ is a \emph{candidate path of type 2} (in $I$) iff there exists $(x,x') \in \S_I$ such that $xPx'$ is a chordless path in $G$. We say that $P$ is a \emph{candidate path} (in $I$) if it is a candidate path of type 1 or 2. In the following, we will simply refer to a candidate path when the instance $I$ is clear from the context.

We first consider the problem of finding a candidate path. This is the aim of the following theorem proved in Section \ref{section4}.

\begin{theorem} \label{thm2} We have an algorithm $\textsc{FindCandidatePath}$ that takes an instance $I = (G, k, X, Y)$ of \dchvdb{}, and in $O(k^2 |V(G)|^4)$ time, either finds $P$ candidate path of type 1, or finds $Q$ candidate path of type 2 with $\lab(Q)$ minimum, or concludes that there is no candidate path.
\end{theorem}

We need to ensure that as long as $G$ is not chordal, we may find a candidate path. We say that the instance $I$ is \emph{well-formed} if: for each $(x,x') \in \P_I - \S_I$, for each path $P$ in $G[V']$ such that $xPx'$ is chordless, we have $y \in Y$ such that $x,x'$ are $y$-connected in $G[W']$ and $\lab(y) \leq \lab(P)$. Essentially, the way we choose a minimum candidate path of type 2 will ensure that our instance is well-formed at each step. We will rely on the following theorem proved in Section \ref{section5}.

\begin{theorem} \label{thm3} Suppose that $I$ is well-formed and that $G[W']$ is chordal. Let $H$ be a hole in $G$ with $|V(H) \cap Z|$ minimum. Let $\F$ be the set of paths $P$ that are $V'$-fragments of $H$ and such that $V(P) \cap Z \neq \emptyset$. Either: (a) $\F = \{P\}$ and $P$ is a candidate path of type 1, or (b) $|\F| \geq 2$, and for $P \in \F$ $\lab$-minimal we have $P$ candidate path of type 2.
\end{theorem}

We need to justify that the recursion depth is at most $3k$. A \emph{model} is a tuple $\mmodel = (X,G,\T)$ with $G = (V,E)$ chordal graph, $X \subseteq V$ and $\T$ clique tree for $G$. Let $Y = V - X$ and suppose that we have $\ppart = (C_1,C_2,\ldots,C_p)$ ordered partition of $Y$. The \emph{length} of $\ppart$ is $|\ppart| = p$. Let $Y_0 = \emptyset$ and for each $i \in [p]$, let $Y_i = \cup_{j \leq i} C_j$. We say that $\ppart$ is a \emph{progressive partition} of $\mmodel$ iff for each $i \in [p]$, there exists $x,x' \in X$ such that $x,x'$ are not $Y_{i-1}$-connected in $G[X \cup Y_{i-1}]$ and $x,x'$ are $Y_i$-connected in $G[X \cup Y_i]$. We will need the following theorem proved in Section \ref{section6}.

\begin{theorem} \label{thm4} Let $\mmodel = (X,G,\T)$ be a model and let $\ppart$ be a progressive partition of $\mmodel$. Then $|\ppart| \leq 2|X|$.
\end{theorem}

\subsection{Statement and proof of the algorithm}

We are now ready to complete the proof of Theorem \ref{thm1} stated in the introduction. We say that the instance $I = (G, k, X, Y)$ is \emph{trivially negative} if $k = 0$ or $G[X \cup Y]$ is not chordal. We solve the \dchvdb{} problem by the following recursive algorithm.

\begin{algorithm}[!]
\caption{$\textsc{SolveDChVD}(I)$}
\label{algo1}
\begin{algorithmic}[1]
\STATE suppose that $I = (G, k, X, Y)$ with $G = (V,E)$ and $k \in \N$
\STATE let $W' = X \cup Y$ and let $Z = V - W'$
\STATE \textbf{if} $k = 0$ or $G[X \cup Y]$ is not chordal \textbf{then} return false
\STATE $P \leftarrow \textsc{FindCandidatePath}(I)$
\STATE \textbf{if} $P = \bot$ \textbf{then} return true
\STATE let $I' = (G, k, X, Y \cup V(P))$
\STATE $r \leftarrow \textsc{SolveDChVD}(I')$
\STATE \textbf{if} $r$ \textbf{then} return true
\FOR{$v \in V(P) \cap Z$}
\STATE let $I' = (G-v, k-1, X, Y)$
\STATE $r \leftarrow \textsc{SolveDChVD}(I')$
\STATE \textbf{if} $r$ \textbf{then} return true
\ENDFOR
\STATE return false
\end{algorithmic}
\end{algorithm}

We prove the correctness of Algorithm \ref{algo1} in the three following lemmas, concluding with the proof of Theorem \ref{thm1}.

\begin{lemma} \label{lem1} Suppose that we consider a call $\textsc{SolveDChVD}(I)$ with $I$ well-formed, and let us consider a subcall $\textsc{SolveDChVD}(I')$ with $I'$ not trivially negative. Then $I'$ is well-formed.
\end{lemma}

\begin{proof} This is clear for a subcall in Line 11. Consider a subcall in Line 7. Let $I' = (G,k,X,Y')$ with $Y' = Y \cup V(P)$, and let $W'' = X \cup Y'$. Note that if $P$ was a candidate path of type 1, we would have $I'$ trivially negative; hence, $P$ is a candidate path of type 2. Consider a pair $(x,x') \in \P_{I'} - \S_{I'}$. We then have $x,x' \in X$ non adjacent in $G$, and $x,x'$ are $Y'$-connected in $G[W'']$. 

Case 1: $x,x'$ are $Y$-connected in $G[W']$. We then have $(x,x') \in \P_I - \S_I$. Consider a path $R$ in $G[V']$ such that $xRx'$ is chordless. Since $I$ is well-formed, we have $y \in Y$ such that $x,x'$ are $y$-connected in $G[W']$ and $\lab(y) \leq \lab(R)$. Since $W' \subseteq W''$ and $Y \subseteq Y'$, we also have $y \in Y'$ such that $x,x'$ are $y$-connected in $G[W'']$ and $\lab(y) \leq \lab(R)$.

Case 2: $x,x'$ are not $Y$-connected in $G[W']$. Since they are $Y'$-connected in $G[W'']$, we find $y \in V(P)$ such that $x,x'$ are $y$-connected in $G[W'']$. Consider a path $R$ in $G[V']$ such that $xRx'$ is chordless. Since $(x,x') \in \S_I$, we have $R$ candidate path of type 2 for instance $I$. By choice of $P$, we have $\lab(P) \leq \lab(R)$, and thus $\lab(y) \leq \lab(R)$. \qed
\end{proof}

\begin{lemma} \label{lem2} Suppose that we consider a call $\textsc{SolveDChVD}(I)$ with $I$ well-formed and such that the call terminates. If $\textsc{SolveDChVD}(I)$ returns true, then $I$ is a positive instance of \dchvdb{}.
\end{lemma}

\begin{proof} We reason by induction on the height the call tree. 

Suppose the call exits in Line 5. By Theorem \ref{thm2}, we know that there is no candidate path for $I$. By Theorem \ref{thm3}, it must follow that $G$ is chordal. Therefore $S = \emptyset$ is a solution for $I$.

Suppose that the call exits in Line 8. We have considered the instance $I' = (G,k,X,Y')$ with $Y' = Y \cup V(P)$. Since $\textsc{SolveDChVD}(I')$ returned true, it is not trivially negative, and by Lemma \ref{lem1} we infer that $I'$ is well-formed. By induction hypothesis we obtain $S$ solution for $I'$. Since $S$ is disjoint from $X \cup Y'$, we also have $S$ disjoint from $X \cup Y$, and thus $S$ is a solution for $I$.

Suppose that the call exits in Line 12. We have considered the instance $I' = (G-v,k-1,X,Y)$. Since $\textsc{SolveDChVD}(I')$ returned true, it is not trivially negative, and by Lemma \ref{lem1} we infer that $I'$ is well-formed. By induction hypothesis we obtain $S$ solution for $I'$. Let $S' = S \cup \{v\}$. Since $v \in Z$, we have $S' \subseteq V$ disjoint from $X \cup Y$ such that $|S'| < k$ and $G - S'$ is chordal. We conclude that $S'$ is a solution for $I$. \qed
\end{proof}

\begin{lemma} \label{lem3} Suppose that we consider a call $\textsc{SolveDChVD}(I)$ with $I$ well-formed and such that the call terminates. If $I$ is a positive instance of \dchvdb{}, then $\textsc{SolveDChVD}(I)$ returns true.
\end{lemma}

\begin{proof} We reason by induction on the height of the call tree.

Let $S$ be a solution for $I$, we then have $S \subseteq V$ disjoint from $X \cup Y$ such that $|S| < k$ and $G - S$ is chordal. Since $|S| < k$, we have $k > 0$. Since $S$ is disjoint from $X \cup Y$, we have $G[X \cup Y]$ induced subgraph of $G - S$ and thus chordal. It follows that the current call cannot exit in Line 3. If the algorithm exits in Line 5, we have the desired conclusion. Suppose now that the algorithm reaches Line 6. 

Case 1: $S$ is disjoint from $V(P)$. Consider the instance $I' = (G, k, X, Y \cup V(P))$. Then $S$ is a solution for $I'$, and thus $I'$ is not trivially negative. By Lemma \ref{lem1}, $I'$ is well-formed. Since $I'$ is positive, by induction hypothesis the subcall $\textsc{SolveDChVD}(I')$ returns true. 

Case 2: $S$ contains a vertex $v \in V(P)$. Consider the instance $I' = (G - v, k - 1, X, Y)$. Then $S - v$ is a solution for $I'$, and thus $I'$ is not trivially negative. By Lemma \ref{lem1}, $I'$ is well-formed. Since $I'$ is positive, by induction hypothesis the subcall $\textsc{SolveDChVD}(I')$ returns true. 
\qed
\end{proof}

We are now ready to complete the proof of Theorem \ref{thm1}. Given an instance $I = (G,k,X,Y)$ of \dchvdb{}, we let $L(I)$ denote the maximum length of a chordless path in $G - X$.

\begin{theorem} (Theorem \ref{thm1} restated) Consider an instance of \dchvdb{} of the form $I = (G,k,X,\emptyset)$ and let $L = L(I)$. Algorithm \ref{algo1} correctly solves instance $I$ in $O((L+1)^{3k} k^2 |V(G)|^4)$ time.
\end{theorem}

\begin{proof} Clearly, $I$ is well-formed. Lemmas \ref{lem2} and \ref{lem3} prove that the algorithm correctly solves the instance $I$, assuming
that it terminates. We now justify the running time of the algorithm. Let $T$ be the tree formed by the recursive calls issued by $\textsc{SolveDChVD}(I)$. Each node $u$ of $T$ is labelled by an instance $I_u$, and we have disjoint sets $R_u, Y_u \subseteq V'$ such that $I_u = (G-R_u,k-|R_u|,X,Y_u)$. Moreover, the set $Y_u$ results by a sequence of additions $C_1,C_2,\ldots,C_l$ to $Y$. We may then define an ordered partition $\P_u = (C_1,C_2,\ldots,C_l)$ of $Y_u$ corresponding to this sequence; by Theorem \ref{thm4}, we always have $|\P_u| \leq 2|X| = 2k$. We then define the measure of a subinstance as $m(I_u) = 3k - |R_u| -|\P_u|$. Note that for an instance $I_u$ we always have $m(I_u) \geq 3k - k - 2k \geq 0$. Moreover, for each child $v$ of $u$ we have $m(I_v) \leq m(I_u) - 1$. Finally, for each subinstance $I_u$ we have $L(I_u) = L$, and thus the node $u$ has at most $L+1$ children. It follows that the recursion tree $T$ has total size at most $(L+1)^{3k}$. Since each recursive call takes $O(k^2 |V(G)|^4)$ by Theorem \ref{thm2}, we conclude that the main call $\textsc{SolveDChVD}(I)$ takes $O((L+1)^{3k} k^2 |V(G)|^4)$ time. \qed
\end{proof}

\section{\label{section4} Proof of Theorem \ref{thm2}}

\subsection{Auxiliary results}

The following lemma provides an efficient algorithm to decide the $w$-connectedness of a pair of vertices $u,v$ in a chordal graph.

\begin{lemma} \label{lem4} Let $G$ be a chordal graph with $n$ vertices, and let $u,v,w \in V(G)$. We may decide in $O(n^3)$ time if $u,v$ are $w$-connected in $G$.
\end{lemma}

\begin{proof} Suppose that $G = (V,E)$ is chordal. We first compute in $O(n^2)$ time a clique tree $\T = (T, \{T_v : v \in V \})$ for $G$ e.g. using the algorithm of \cite{RTL76}. We may suppose that $T = (W,F)$ with $|W| \leq n$. For each edge $xy$ of $T$, we compute the separation $S^{\T}_{xy} = (A,S,B)$. If $w \in S$, we test whether there is an $(u,w)$-path in $G[A \cup \{w\}]$ and a $(v,w)$-path in $G[B \cup \{w\}]$, and we accept if this is the case. If no such edge $xy$ is found, we reject. The correctness of the algorithm is easy to see, and its time complexity is $O(n^3)$ since $|F| \leq n$, and since each inspection of an edge of $F$ takes $O(n^2)$ time. \qed
\end{proof}

We now define some specific set of paths in a graph.
Let $G = (V,E)$ be a graph. Consider three sets $A,B,C \subseteq V$. We denote by $\Pi^1_G(A)$ the set of paths of length 1 in $G$ that have the form $P = z$ with $z \in A$. We denote by $\Pi^2_G(A,B)$ the set of paths of length 2 in $G$ that have the form $P = z_1 z_2$ with $z_1 \in A, z_2 \in B$. We denote by $\Pi^{\geq 3}_G(A,B,C)$ the set of chordless paths of length at least 3 in $G$ that have the form $P = z_1 z_2 \ldots z_L$ with $z_1 \in A, z_L \in B$ and for each $i$ ($1 < i < L$) $z_i \in C$.

Note that the set $\Pi^{\geq 3}_G(A,B,C)$ has exponential size in general. However, we will only need to find a minimum-weight path in the set, which can be done efficiently as follows.

\begin{lemma} \label{lem5} Consider a graph $G$ with $n$ vertices, a weight function $w : V(G) \rightarrow \N$, and three sets $A,B,C \subseteq V(G)$. Given a path $P$ in $G$, let $w(P) = \max_{v \in V(P)} w(v)$. In $O(n^3)$ time we can compute a path $P \in \Pi^{\geq 3}_G(A,B,C)$ with $w(P)$ minimum, or report that no such path exists.
\end{lemma}

\begin{proof} Fix $a \in A$. Let $B' = B - N_G[a]$ and let $S = \{a\} \cup B' \cup C$. We construct the weighted graph $\H = (G',w')$ where $G' = G[S]$ and $w' = w | S$. Using Dijkstra algorithm \cite{D59} adapted to the $(\max,\min)$-setting, we compute the distances $d_{\H}(a,v)$ for $v \in S$ in $O(n^2)$ time. We choose $b \in B'$ such that $d_{\H}(a,b)$ is minimum, and this gives the weight of a minimum path in $\Pi^{\geq 3}_G(\{a\},B,C)$. By iterating over the vertices $a \in A$ and by selecting the vertex that gives the minimum result, we find the weight of a minimum path in $\Pi^{\geq 3}_G(A,B,C)$ in $O(n^3)$ time. Using appropriate chaining, we can recover such a minimum path in $O(n^2)$ time. \qed
\end{proof}

\subsection{Proof of the main result}

Consider an instance $I = (G,k,X,Y)$ of \dchvdb{}. As above, we have $G = (V,E)$, $W' = X \cup Y$, $Z = V - W'$ and $V' = Y \cup Z$. We define the following notions.
\begin{itemize}
\item A \emph{1-anchor} in $I$ is a tuple $(x)$ with $x \in X$.
\item A \emph{2-anchor} in $I$ is a tuple $(x,x')$ with $x,x' \in X$ adjacent in $G$.
\item A \emph{3-anchor} in $I$ is a tuple $(x,x',y)$ with $x,x' \in X$ nonadjacent in $G$, and $y \in W'$ such that $x,x'$ are $y$-connected in $G[W']$.
\item A \emph{4-anchor} in $I$ is a tuple $(x,x')$ with $x,x' \in X$ nonadjacent in $G$, such that $x,x'$ are not $Y$-connected in $G[W']$.
\end{itemize}
For $i \in [4]$, we define $\S^i_I$ as the set of $i$-anchors in $I$. Note that $\S^4_I$ coincides with the set $\S_I$ considered in Section \ref{section3}. For each $t \in \S^i_I$, we define a set of paths $\P^i_{I,t}$ as follows.
\begin{itemize}
\item Let $t = (x)$ be a 1-anchor in $I$. Let $A_t = \{ z \in V' : N_G(z) \cap \{x\} = \{x\} \}$ and $B_t = \{ z \in V' : N_G(z) \cap \{x\} = \emptyset \}$. We define $\P^1_{I,t} = \Pi^{\geq 3}_G(A_t,A_t,B_t)$.
\item Let $t = (x,x')$ be a 2-anchor in $I$. Let $A_t = \{ z \in V' : N_G(z) \cap \{x,x'\} = \{x\} \}$, $B_t = \{ z \in V' : N_G(z) \cap \{x,x'\} = \{x'\} \}$ and $C_t = \{ z \in V' : N_G(z) \cap \{x,x'\} = \emptyset \}$. We define $\P^2_{I,t} = \Pi^2_G(A_t,B_t) \cup \Pi^{\geq 3}_G(A_t,B_t,C_t)$.
\item Let $t = (x,x',y)$ be a 3-anchor in $I$. Let $A_t = \{ z \in V' : N_G(z) \cap \{x,x',y\} = \{x\}$, $B_t = \{ z \in V' : N_G(z) \cap \{x,x',y\} = \{x'\} \}$, $C_t = \{ z \in V' : N_G(z) \cap \{x,x',y\} = \emptyset \}$ and $D_t = \{ z \in V' : N_G(z) \cap \{x,x',y\} = \{x,x'\} \}$. We define $\P^3_{I,t} = \Pi^1_G(D_t) \cup \Pi^2_G(A_t,B_t) \cup \Pi^{\geq 3}_G(A_t,B_t,C_t)$.
\item Let $t = (x,x')$ be a 4-anchor in $I$. Let $A_t = \{ z \in V' : N_G(z) \cap \{x,x'\} = \{x\} \}$, $B_t = \{ z \in V' : N_G(z) \cap \{x,x'\} = \{x'\} \}$, $C_t = \{ z \in V' : N_G(z) \cap \{x,x'\} = \emptyset \}$  and $D_t = \{ z \in V' : N_G(z) \cap \{x,x'\} = \{x,x'\} \}$. We define $\P^4_{I,t} = \Pi^1_G(D_t) \cup \Pi^2_G(A_t,B_t) \cup \Pi^{\geq 3}_G(A_t,B_t,C_t)$.
\end{itemize}

The following two lemmas prove that, in order to find a candidate path, we may restrict ourselves to the sets $\P^i_{I,t}$.

\begin{lemma} \label{lem6} Let $P$ be a chordless path in $G[V']$.
\begin{enumerate}
\item If $P$ is a candidate path of type 1 in $I$, then $P \in \P^i_{I,t}$ for some $i \in [3]$ and $t \in \S^i_I$.
\item If $P \in \P^i_{I,t}$ for some $i \in [3]$ and $t \in \S^i_I$, then we can find $P'$ candidate path of type 1 in $I$.
\end{enumerate}
\end{lemma}

\begin{proof} Point 1. Suppose that $P$ is a candidate path of type 1 in $I$. We then find $H$ hole in $G$ such that (a) $P$ is a $V'$-fragment of $H$, (b) for each $Q$ $V'$-fragment of $H$ distinct of $P$, we have $V(Q) \subseteq Y$. Suppose that $H = z_1 z_2 \ldots z_L z_1$ and suppose w.l.o.g. that $P = z_1 z_2 \ldots z_p$. We then have $z_L \in X$. If $p = L-1$, we obtain that $t = (z_L) \in \S^1_I$ and that $P \in \Pi^{\geq 3}_G(A_t, A_t, B_t)$, hence $P \in \P^1_{I,t}$. If $p = L-2$, we obtain that $t = (z_L,z_{L-1}) \in \S^2_I$ and that $P \in \Pi^2_G(A_t,B_t) \cup \Pi^{\geq 3}_G(A_t,B_t,C_t)$, hence $P \in \P^2_{I,t}$. Suppose now that $p < L-2$. Let $x = z_L, x' = z_{p+1}$ and $y = z_{p+2}$. Observe that the path $z_{p+1} z_{p+2} \ldots z_L$ contains $y$ and is included in $W'$. It follows that $x,x'$ are $y$-connected in $G[W']$, and thus $t = (x,x',y)$ is in $\S^3_I$. Moreover, we have $P \in \Pi^1_G(D_t) \cup \Pi^2_G(A_t,B_t) \cup \Pi^{\geq 3}_G(A_t,B_t,C_t)$ and thus $P \in \P^3_{I,t}$.

Point 2. Suppose that $P \in \P^i_{I,t}$ for some $i \in [3]$ and $t \in \S^i_I$. If $i = 1$, we have $t = (x)$ with $x \in X'$, and since $P \in \P^1_{I,t}$ we have $H = xPx$ hole in $G$ and thus $P$ is a candidate path of type 1 in $I$. If $i = 2$, we have $t = (x,x')$ with $x,x' \in X'$ adjacent in $G$, and since $P \in \P^2_{I,t}$ we have $H = xPx'x$ hole in $G$ and thus $P$ is a candidate path of type 2 in $I$. Suppose now that $i = 3$. We then have $t = (x,x',y)$ with $x,x' \in X$ nonadjacent in $G$ and $y \in W'$ such that $x,x'$ are $y$-connected in $G[W']$. We  find $Q$ chordless $(x,x')$-path in $G[W']$ containing $y$. Suppose that $P = z_1 z_2 \ldots z_p$ and suppose that $Q = z'_1 z'_2 \ldots z'_q$. We then have $z'_1 = x, z'_q = x'$ and $z'_r = y$ for some $r$ ($1 < r < q$). Let $J$ be the set of pairs $(\alpha,\beta)$ ($1 \leq \alpha \leq p, 1 \leq \beta \leq r$) such that $z_{\alpha} z'_{\beta} \in E$; observe that $J$ contains $(1,1)$. Let us choose $(\alpha,\beta) \in J$ with $\alpha+\beta$ maximum. Let $J'$ be the set of pairs $(\gamma,\delta)$ ($\alpha \leq \gamma \leq p, r \leq \delta \leq q$) such that $z_{\gamma} z'_{\delta} \in E$; observe that $J'$ contains $(p,q)$. Choose $(\gamma,\delta) \in J'$ with $\gamma+\delta$ minimum. Since the vertices of $P$ are not adjacent to $y$, we have $\beta < r < \delta$. Let $P' = P[z_{\alpha}, z_{\gamma}]$ and $Q' = Q[z'_{\beta}, z'_{\delta}]$, then $E(P') \cup E(Q')$ form a hole $H'$ in $G$. Hence, $P'$ is included in a $V'$-fragment $R$ of $H'$, and for each $R'$ $V'$-fragment of $H'$ distinct of $R$ we have $V(R') \subseteq V(Q') \subseteq Y$. We conclude that $P'$ is a candidate path of type 1 in $I$.
 \qed
\end{proof}

\begin{lemma} \label{lem7} Let $P$ be a chordless path in $G[V']$.
\begin{enumerate}
\item If $P$ is a candidate path of type 2 in $I$, then $P \in \P^4_{I,t}$ for some $t \in \S^4_I$.
\item If $P \in \P^4_{I,t}$ for some $t \in \S^4_I$, then $P$ is a candidate path of type 2 in $I$.
\end{enumerate}
\end{lemma}

\begin{proof} Point 1. Suppose that $P$ is a candidate path of type 2 in $I$. We then find $(x,x') \in \S_I$ such that $xPx'$ is a chordless path in $G[V']$. Let $t = (x,x')$, we then have $t \in \S^4_I$. Suppose that $P = z_1 z_2 \ldots z_p$. If $p = 1$, we have $z_1 \in D_t$. If $p > 1$, we have $z_1 \in A_t, z_p \in B_t$, and $z_i \in C_t$ ($1 < i < p$). It follows that: $P \in \Pi^1_G(D_t) \cup \Pi^2_G(A_t,B_t) \cup \Pi^{\geq 3}_G(A_t,B_t,C_t)$. We conclude that $P \in \P^4_{I,t}$.

Point 2. Suppose that $P \in \P^4_{I,t}$ for some $t = (x,x') \in \S^4_I$. Suppose that $P = z_1 z_2 \ldots z_p$. If $P \in \Pi^1_G(D_t)$, then $z_1 \in D_t$ and $xPx'$ is a chordless path in $G$. If $P \in \Pi^2_G(A_t,B_t)$, then $z_1 \in A_t, z_2 \in B_t$, hence $xPx'$ is a chordless path in $G$. If $P \in \Pi^{\geq 3}_G(A_t,B_t,C_t)$, then $z_1 \in A_t, z_p \in B_t$, and $z_i \in C_t$ ($1 < i < p$), hence $xPx'$ is a chordless path in $G$. Since $(x,x') \in \S_I$, we conclude that $P$ is a candidate path of type 2 in $I$.
 \qed
\end{proof}

We are now ready to complete the proof of Theorem \ref{thm2}.

\begin{theorem} (Theorem \ref{thm2} restated) We have an algorithm \textsc{FindCandidatePath} that takes an instance $I = (G,k,X,Y)$ of \dchvdb{}, and in $O(k^2 |V(G)|^4)$ time, either finds $P$ candidate path of type 1, or finds $Q$ candidate path of type 2 with $\lab(Q)$ minimum, or concludes that there is no candidate path.
\end{theorem}

\begin{proof} Let $n = |V(G)|$.

The algorithm first constructs the set of anchors. We can test in constant time whether a tuple is a 1-anchor or a 2-anchor. To decide if a tuple $(x,x',y)$ is a 3-anchor, we use Lemma \ref{lem4} to decide in $O(n^3)$ time if $x,x'$ are $y$-connected in $G[W']$. To decide if a tuple $(x,x')$ is a 4-anchor, we use Lemma \ref{lem4} to decide in $O(n^4)$ time are $Y$-connected in $G[W']$. Hence, we can list the anchors in $O(k^2 n^4)$ time.

Suppose now that the set of anchors is available. To find a candidate path of type 1, by Lemma \ref{lem6} it suffices to test if there is some $i \in [3]$ and some $t \in \S^i_I$ such that $\P^i_{I,t}$ is non-empty. Similarly, to find a minimum candidate path of type 2, by Lemma \ref{lem7} it suffices to compute the minimum of $\lab(P)$ over each tuple $t \in \S^4_I$ and each path $P \in \P^4_{I,t}$. By Lemma \ref{lem5}, for a given $t$ the computation can be done in $O(n^3)$ time. As there are $O(k^2 n)$ 3-anchors and $O(k^2)$ 4-anchors, we may find the desired result in $O(k^2 n^4)$ time. \qed
\end{proof}

\section{\label{section5} Proof of Theorem \ref{thm3}}

We will need the following standard property of chordal graphs.

\begin{lemma} \label{lem8} Suppose that $G$ is chordal. Let $P = z_1 z_2 \ldots z_L$ a chordless path in $G$ with $L \geq 3$ and let $R \subseteq V(G)$ such that (a) $G[R]$ is connected, (b) $z_1, z_L \in N_G[R]$. Then for each $i \in [L]$ we have $z_i \in N_G[R]$.
\end{lemma}

\begin{proof} By way of contradiction, suppose that $z_i \notin N_G[R]$ for some $i$ ($1 < i < L$). Let $J = \{ p : 1 \leq p < i \text{ and } z_i \in N_G[R] \}$ and $J' = \{ p : i < p \leq L \text{ and } z_i \in N_G[R] \}$. We have $J \neq \emptyset$ since $z_1 \in N_G[R]$, and we have $J' \neq \emptyset$ since $z_L \in N_G[R]$. Choose $p$ maximum element of $J$, and $q$ minimum element of $J'$. Since $G[R]$ is connected, we find $Q$ chordless $(z_p,z_q)$-path which internal vertices are in $R$. Then $Q z_{q-1} \ldots z_p$ is a hole in $G$, contradiction.
\qed
\end{proof}

\begin{figure}[hbt]
\center
\includegraphics[scale=0.5]{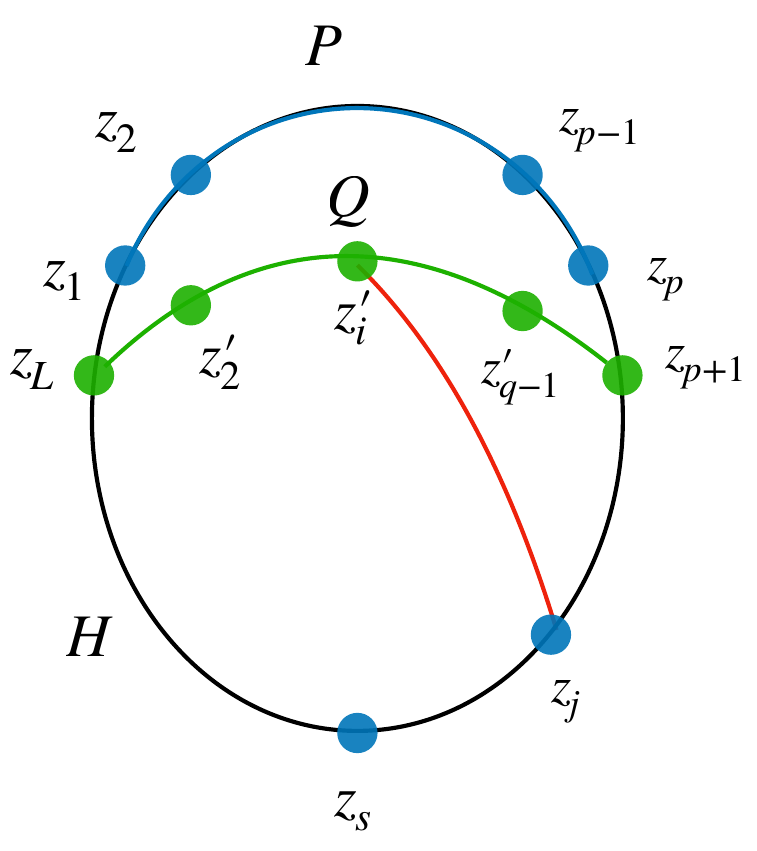}
\caption{\label{fig1}Illustrating the situation in Claim 2.}
\end{figure}

We are now in a position to prove Theorem \ref{thm3}.

\begin{theorem} (Theorem \ref{thm3} restated) Suppose that $I$ is well-formed and that $G[W']$ is chordal. Let $H$ be a hole in $G$ with $|V(H) \cap Z|$ minimum. Let $\F$ be the set of paths $P$ that are $V'$-fragments of $H$ and such that $V(P) \cap Z \neq \emptyset$. Either: (a) $\F = \{P\}$ and $P$ is a candidate path of type 1, or (b) $|\F| \geq 2$, and for $P \in \F$ $\lab$-minimal we have $P$ candidate path of type 2.
\end{theorem}

\begin{proof} Using the perfect elimination ordering $\sigma$, we may construct a clique tree for $G[V']$ of the form $\T = (T, \{T_v : v \in V'\})$. Suppose that $T = (W,F)$, and that $T$ is rooted at some node $r_T \in W$. Given two nodes $u,v \in W$, we denote $u \leq_T v$ if $v$ is closer to $r_T$ than $u$. Given a connected subset $S$ of $T$, we let $r(S)$ denote the element of $S$ maximal for $\leq_T$. Given $v \in V'$, we let $r(v) = r(T_v)$. We may assume that: given $u,v \in V'$, if $r(u) \leq_T r(v)$ then $\lab(u) \leq \lab(v)$.

Since $G[W']$ is chordal, the hole $H$ must intersect $Z$, hence $\F$ is non-empty. Clearly, if $\F = \{P\}$ then $P$ is a candidate path of type 1. Suppose now that $|\F| \geq 2$, and let $P \in \F$ $\lab$-minimal. Suppose that $H = z_1 z_2 \ldots z_L z_1$ and that $P = z_1 z_2 \ldots z_p$ with $p < L$. Consider $x = z_L$ and $x' = z_{p+1}$, we then have $x,x' \in X$ nonadjacent. Consider $P' \in \F$ distinct from $P$, then $V(P') \cap Z$ contains some vertex $z_s$ with $p+1 < s < L$.

If $P$ is a candidate path of type 2, we're done. Suppose the contrary. Since $xPx'$ is a chordless path in $G$, we have $(x,x') \notin \S_I$. As $I$ is well-formed, we have $y \in Y$ such that $x,x'$ are $y$-connected in $G[W']$ and $\lab(y) \leq \lab(P)$. We then find $Q$ chordless $(x,x')$-path in $G[W']$ containing $y$. Let us suppose that $Q = z'_1 z'_2 \ldots z'_q$, with $z'_1 = x, z'_q = x'$ and $z'_r = y$ for some $r$ ($1 < i < q$). 

\textit{Claim 1.} $r(y) \leq_T r(P)$.

\textit{Proof.} If we had $r(P) <_T r(y)$ it would follow that $\lab(P) < \lab(y)$, contradiction. Thus, we have either (a) $r(y) \leq_T r(P)$ or (b) $r(y)$ and $r(P)$ are incomparable under $\leq_T$. Towards a contradiction, suppose that we are in case (b). For each $u \in V(P)$, we have $r(u) \leq_T r(P)$, which implies that $r(u)$ and $r(y)$ are incomparable under $\leq_T$; hence, $y \notin N_G[V(P)]$. Consider $G' = G[V(P) \cup V(Q)]$. Since $|V(G') \cap Z| < |V(H) \cap Z|$, we have $G'$ chordal by choice of $H$. Consider $R = V(P)$. We then have $G'[R]$ connected and $z'_1,z'_q\in N_G[R]$. By Lemma \ref{lem8}, we have $z'_r \in N_G[R]$. But we have seen that $y \notin N_G[R]$, contradiction. \qed

We infer from Claim 1 that $y,z_s$ are distinct and not adjacent in $G$. Indeed, suppose by contradiction that $y = z_s$ or $y z_s \in E$. By minimality of $P$, we have either $r(P) <_T r(P')$ or $r(P), r(P')$ incomparable under $\leq_T$. Since $r(y) \leq_T r(P)$, we must have $z_s$ adjacent to $r(P)$, contradiction.

\textit{Claim 2.} (a) For each $i,j$ ($2 \leq i < q-1, p+1 \leq j < s$), we have $z'_i z_j \notin E$; (b) For each $i,j$ ($2 < i \leq q-1, s < j \leq L$), we have $z'_i z_j \notin E$.

\textit{Proof.} We only prove (a) as (b) follows by symmetry. Towards a contradiction, assume that $z'_i z_j \in E$ for some $i,j$ ($2 \leq i < q-1, p+1 \leq j < s$). Refer to Figure \ref{fig1} for an illustration. Consider $G' = G[\{z_L,z_1,z_2,\ldots,z_{j}\} \cup V(Q)\}]$. Since $|V(G') \cap Z| < |V(H) \cap Z|$, we have $G'$ chordal by choice of $H$. Consider $P' = z_L z_1 \ldots z_j$ and $R = \{z'_2,z'_3,\ldots,z'_i\}$, then $P'$ is a chordless path in $G'$ and $R$ is a connected subset of $G'$. Since $z_L, z_j \in N_{G'}[R]$ and since $G'$ is chordal, it follows from Lemma \ref{lem8} that $z_{p+1} \in N_{G'}[R]$. But we have $z_{p+1} = z'_q$ distinct from and nonadjacent to $z'_2,\ldots,z'_i$, contradiction. 
\qed
 
Recall that we have $z'_r = y$. Let $J$ be the set of pairs $(\alpha,\beta)$ such that $2 \leq \alpha \leq r$, $s \leq \beta \leq L$ and $z'_{\alpha} z_{\beta} \in E$. Let $J'$ be the set of pairs $(\gamma,\delta)$ such that $r \leq \gamma \leq q-1$, $p+1 \leq \delta \leq s$ and $z'_{\gamma} z_{\delta} \in E$. Observe that $J$ contains $(2,L)$ and that $J'$ contains $(q-1,p+1)$. Let us choose $(\alpha,\beta) \in J$ with $\beta-\alpha$ minimum, and let us choose $(\gamma,\delta) \in J'$ with $\delta-\gamma$ maximum. Let us consider $P' = P[z_{\delta},z_{\beta}]$ and $Q' = Q[z'_{\alpha},z'_{\gamma}]$. 

\textit{Claim 3.} $E(P') \cup E(Q')$ form a hole in $G$.

\textit{Proof.} Let $C = E(P') \cup E(Q')$. We first show that $|P'| + |Q'| \geq 4$. By contradiction, suppose that $|P'| + |Q'| \leq 3$. We have the following cases. Suppose first that $|Q'| = 1$ and $|P'| \leq 2$. We then have $\alpha = \gamma = r$ and thus $Q' = y$. But $z_s$ is an endpoint of $P'$ and must be adjacent to $y$, contradiction. Suppose now that $|Q'| = 2$ and $|P'| = 1$. We then have $\beta = \delta = s$ and thus $P' = z_s$. But $y$ is an endpoint of $Q'$ and must be adjacent to $z_s$, contradiction.

Suppose that $C$ fails to be a chordless cycle. Let $S$ be the set of pairs $(i,j)$ with $\alpha \leq i \leq \gamma$ and $\delta \leq j \leq \beta$ such that $(i,j) \neq (\alpha,\beta), (\gamma,\delta)$. Let $S_1$ be the set of pairs $(i,j) \in S$ such that $z'_i = z_j$, and let $S_2$ be the set of pairs $(i,j) \in S$ such that $z'_i z_j \in E$. Since $C$ fails to be a chordless cycle, we have $S_1 \cup S_2 \neq \emptyset$. Suppose that we have $(i,j) \in S_1$. If either (a) $|P'| = 1$ and $|Q'| = 3$ or (b) $|P'| = 3$ and $|Q'| = 1$, we must have $i = r$, $j = s$, but $z'_r$ and $z_s$ are distinct, contradiction. Otherwise, we conclude that either one of $(i-1,j), (i+1,j)$ is in $S_2$, or one of $(i,j-1), (i,j+1)$ is in $S_2$. 

Thus, we may assume that $S_2 \neq \emptyset$. 
Consider $(i,j) \in S_2$. Let us assume first that $i = r$. If $j \geq s$, we have $(i,j) \in J$ with $j-i < \beta-\alpha$, contradicting the choice of $(\alpha,\beta)$. If $j \leq s$, we have $(i,j) \in J'$ with $j-i > \delta-\gamma$, contradicting the choice of $(\gamma,\delta)$.
Let us assume now that $i < r$. We cannot have $p+1 \leq j < s$ by Claim 2. Thus, we have $s \leq j \leq L$. But then $(i,j) \in J$ with $j-i < \beta-\alpha$, contradicting the choice of $(\alpha,\beta)$. Finally, the case $i > r$ follows by symmetry.
\qed

By Claim 3, we obtain that $E(P') \cup E(Q')$ form a hole $H'$ in $G$. Since $V(H') \subseteq (V(H) - V(P)) \cup V(Q)$, $V(P) \cap Z \neq \emptyset$ and $V(Q) \cap Z = \emptyset$, we have $|V(H') \cap Z| < |V(H) \cap Z|$, contradicting the choice of $H$.
\qed
\end{proof}

\section{\label{section6} Proof of Theorem \ref{thm4}}

\subsection{Auxiliary results}

This section consists of three lemmas that will prove useful in order to establish Theorem \ref{thm4}.

\begin{lemma} \label{lem9} Let $G = (V,E)$ be a chordal graph with clique tree $\T = (T,\{T_v : v \in V\})$. Suppose that $T = (W,F)$ and $K$ is a connected subset of $W$. Let $Z = \{ v \in V : V(T_v) \cap K \neq \emptyset \}$. Consider a chordless path $P = z_1 z_2 \ldots z_p$ in $G$ with $V(P) \cap Z = \{z_p\}$. There exists an edge $e = xy \in F$ with $x \in K, y \notin K$ such that $S^{\T}_{xy} = (A,S,B)$ with $z_p \in S$ and $z_i \in B$ ($1 \leq i < p$).
\end{lemma}

\begin{proof} Consider $U = \cup_{i \in [p-1]} V(T_{z_i})$. Since $P$ is a path, $U$ is a connected subset of $T$. If we had $U \cap K \neq \emptyset$, we would find $i \in [p-1]$ such that $V(T_{z_i}) \cap K \neq \emptyset$ and thus $z_i \in Z$, contradiction. Thus, $U$ and $K$ are disjoint connected subsets of $T$. Choose $Q$ $(u,v)$-path in $T$ with $u \in K, v \in U$ and $Q$ of minimum length. Suppose that $Q = z'_1 z'_2 \ldots z'_q$. Let $x = z'_1, y = z'_2$, and consider $e = x y$ edge in $T$. By choice of $Q$, we have $x \in K$ and $y \notin K$. Consider $S^{\T}_{xy} = (A,S,B)$. Let $C$, resp. $C'$, be the connected component of $T - e$ containing $x$, resp. $y$. By choice of $Q$, we have $K \subseteq C$ and $U \subseteq C'$. It follows that $z_i \in B$ for each $i$ ($1 \leq i < p$). Let us prove that $z_p \in S$. Since $z_p$ is adjacent to $z_{p-1}$, we find $x \in V(T_{z_p}) \cap V(T_{z_{p-1}})$. Since $z_p \in Z$, we find a vertex $y \in V(T_{z_p}) \cap K$. Since $V(T_{z_{p-1}}) \subseteq C'$, we have $x \in C'$. Since $K \subseteq C$, we have $y \in C$. Since $T_{z_p}$ contains $x$ and $y$, it contains the edge $e$, and thus $z_p \in S$. \qed
\end{proof}

\begin{lemma} \label{lem10} Let $G = (V,E)$ be a graph, let $u,v,w \in V$ and let $P$ be a chordless $(u,v)$-path in $G$ containing $w$ as internal vertex. Suppose that there is $S \subseteq V$ such that $G[S]$ is connected and $N_G[w] \subseteq N_G[S]$. We then find $P'$ chordless $(u,v)$-path in $G$ such that (a) $V(P') \subseteq V(P) - w + S$, (b) $I(P') \cap S \neq \emptyset$.
\end{lemma}

\begin{proof} Suppose that $P = z_1 z_2 \ldots z_L$ with $z_i = w$ for some $i$ ($1 < i < L$). Since $z_{i-1}, z_{i+1} \in N_G(w)$, they are in $N_G[S]$. Consider $p \leq i-1$ minimum such that $z_p \in N_G[S]$, and consider $q \geq i+1$ maximum such that $z_q \in N_G[S]$. Since $G[S]$ is connected, we find $Q$ chordless $(z_p,z_q)$-path in $G$ with $I(Q) \subseteq S$. We conclude that $P' = z_1 z_2 \ldots z_{p-1} Q z_{q+1} \ldots z_L$ is a chordless $(u,v)$-path in $G$. Clearly, we have $V(P') \subseteq V(P) - w + S$, and since $Q$ has length at least 3 we have $I(P') \cap S \neq \emptyset$.  \qed
\end{proof}

\begin{lemma} \label{lem11} Let $G = (V,E)$ be a chordal graph and let $\T = (T,\{T_v : v \in V \})$ be a clique tree for $G$. Let $w \in V$ and $S \subseteq V$. Suppose that $(P_1)$ for each $z \in S$, $E(T_z) \cap E(T_w) \neq \emptyset$ and $(P_2)$ $E(T_w) \subseteq \cup_{z \in S} E(T_z)$. We then have: (a) $G[S]$ is connected and (b) $N_G[w] \subseteq N_G[S]$.
\end{lemma}

\begin{proof} Let us prove Point (a). Consider $U = \cup_{z \in S} T_z$. We claim that $U$ forms a subtree of $T$. Suppose not, we then find an edge $e = xy$ of $T$ that separates $U$ in two disconnected subsets. Let us assume that we have a subtree $T_u$ on the $x$-side of $e$ and a subtree $T_v$ on the $y$-side of $e$. By $(P_2)$, $T_w$ cannot contain the edge $e$, and we may assume that $T_w$ is on the $y$-side of $e$. But then we have $E(T_u) \cap E(T_w) = \emptyset$, which contradicts $(P_1)$. Thus, $U$ is a subtree of $T$, and  we conclude that $G[W]$ is connected.

Let us prove Point (b). Consider $u \in N_G[w]$, we then have $V(T_u) \cap V(T_w) \neq \emptyset$. Consider $x$ in the intersection. By $(P_2)$, we have $x \in V(T_z)$ for some $z \in S$. We then have $V(T_u) \cap V(T_z) \neq \emptyset$ and thus $u \in N_G[S]$. 
\qed
\end{proof}

\subsection{Proof of the main result}

In this section, we consider a model $\mmodel = (X,G,\T)$ where $G = (V,E)$, $\T = (T, \{T_v : v \in V\})$ and $T = (W,F)$. We let $Y = V - X$. We suppose that $T$ is supposed to be rooted at some node $r_T$. Given $u \in W$, we denote $p(u)$ its parent in $T$, or $\bot$ if $u = r_T$. 

For each node $x \in W$, let $D^{\T}_x = C^{\T}_x \cap \xset$. We define the digraph $D_{\mmodel} = (W,A_{\mmodel})$ where $A_{\mmodel}$ contains the arc $(x,y)$ whenever $xy \in F$ and $D^{\T}_x \subseteq D^{\T}_y$. A \emph{component} of $D_{\mmodel}$ is an inclusionwise maximal connected subset $K \subseteq W$ such that $D^{\T}_x = D^{\T}_y$ whenever $x,y \in K$. A \emph{maximal component} of $D_{\mmodel}$ is a component $K$ of $D_{\mmodel}$ such that there is no arc $(x,y) \in A_{\mmodel}$ with $x \in K, y \notin K$. A \emph{reducible edge} in $\mmodel$ is an edge $uv \in F$ such that (a) $v = p(u)$, (b) $(u,v) \in A_{\mmodel}$, (c) for each $w$ child of $u$ in $T$, we have $(u,w) \notin A_{\mmodel}$. 

Consider a connected subset $S$ of $W$. The \emph{contraction of $S$ in $\mmodel$} produces the model $\mmodel' = (X,G',\T')$ defined as follows. On the one hand, we have $\T' = (T', \{T'_v : v \in V(G)\})$ where $T'$ (resp. $T'_v$) is obtained by contracting the vertices of $S$ in $T$ (resp. $T_v$),  On the other hand, we have $V(G') = V(G)$ and $E(G') = E(G) \cup \{ uv : u,v \in V(G), V(T_u) \cap S \neq \emptyset \text{ and } V(T_v) \cap S \neq \emptyset \}$. We denote by $\mmodel / S$ the result of contracting $S$ in $\mmodel$. We say that $\mmodel'$ is a \emph{minor} of $\mmodel$ iff there exists $S$ such that $\mmodel' = \mmodel / S$. 

In the following lemmas, we assume that we are given an ordered partition $\ppart = (C_1,\ldots,C_p)$ of $Y$ that is a progressive partition of $\mmodel$. We let $\yset_0 = \emptyset$ and $\yset_i = \cup_{j \leq i} C_j$ for $i \in [p]$. We let $W'_i = \xset \cup \yset_i$ for each $i$ ($0 \leq i \leq p$). The subsequent lemma addresses the case when $D_{\mmodel}$ is not acyclic. 

\begin{lemma} \label{lem12} 
Suppose that $K$ is a maximal component of $D_{\mmodel}$, and let $\mmodel' = \mmodel / K$. Then $\ppart$ is a progressive partition of $\mmodel'$.
\end{lemma}

\begin{proof} Suppose that $\mmodel' = (X,G',\T')$. Let $Z = \{ v \in V : V(T_v) \cap K \neq \emptyset \}$. Observe that for each $v \in Z \cap \xset$, we have $K \subseteq V(T_v)$. We then have $E(G') = \{ uv : uv \in E \text{ or } u,v \in Z \}$. We show that $\ppart$ is a progressive partition of $\mmodel'$. Consider $i \in [p]$. We then find $u,v \in \xset$ that are not $\yset_{i-1}$-connected in $G[W'_{i-1}]$ and are $\yset_i$-connected in $G[W'_i]$.

First point: $u,v$ are $\yset_i$-connected in $G'[W'_i]$. Since $u,v$ are $\yset_i$-connected in $G[W'_i]$, we find $P$ chordless $(u,v)$-path in $G$ such that $V(P) \subseteq W'_i$ and $V(P) \cap \yset_i \neq \emptyset$. Suppose that $P = z_1 z_2 \ldots z_L$ with $z_1, z_L \in \xset$. If $P$ is chordless in $G'$, we have the desired conclusion. Suppose now that $P$ has a chord in $G'$. We may assume that the chord is $z_p z_q$ with $q-p$ maximum. Hence $P' = z_1 z_2 \ldots z_p z_q \ldots z_L$ is a chordless path in $G'$, and we have $z_p,z_q \in Z$. If we had $z_p \in \xset$, we would have $K \subseteq V(T_{z_p})$ and $V(T_{z_q}) \cap K \neq \emptyset$, hence $z_p z_q \in E$, contradiction. Thus, $z_p$ is in $\yset_i$. We deduce that $P'$ is a chordless $(u,v)$-path in $G'$ such that $V(P') \subseteq W'_i$ and $V(P') \cap \yset_i \neq \emptyset$. We obtain that $u,v$ are $\yset_i$-connected in $G'[W'_i]$.

Second point: $u,v$ are not $\yset_{i-1}$-connected in $G'[W'_{i-1}]$. By way of contradiction, suppose that we have $P$ chordless $(u,v)$-path in $G'$ such that $V(P) \subseteq W'_{i-1}$ and $V(P) \cap \yset_{i-1} \neq \emptyset$. Suppose that $P = z_1 z_2 \ldots z_L$. If $P$ is a path in $G$, we obtain that $u,v$ are $\yset_{i-1}$-connected in $G[W'_{i-1}]$, contradiction. Suppose now that the edge $z_p z_{p+1}$ is not in $G$. We then have $z_p, z_{p+1} \in Z$. If we had $z_p \in \xset$ or $z_{p+1} \in \xset$, we would obtain that $z_p z_{p+1} \in E$, contradiction. Thus, we have $z_p, z_{p+1} \in \yset_{i-1}$. Moreover, for each $j < p$ or $j > p+1$ we have $z_j \notin Z$, since otherwise $z_j$ would be adjacent to both $z_p, z_{p+1}$ in $G'$. By Lemma \ref{lem9}, we find an edge $e = xy$ in $T$ such that $x \in K, y \notin K$ and $S^{\T}_{xy} = (A,S,B)$ with $z_p \in S$ and $z_i \in B$ ($i < p$). By choice of $K$, we have $(x,y) \notin A_{\mmodel}$ and thus $D^{\T}_{x} \nsubseteq D^{\T}_{y}$. Therefore, we find a vertex $w \in D^{\T}_{x} - D^{\T}_{y}$. We then have $z_p w \in E$ and $z_j w \notin E$ ($j < p$). Moreover, since $w \in \xset$ we have $K \subseteq V(T_w)$ and thus $z_{p+1} w \in E$. Consider $q$ ($q \geq p+1$) maximum such that $z_q w \in E$. We obtain $P' = z_1 z_2 \ldots z_p w z_q \ldots z_L$ chordless $(u,v)$-path in $G$ such that $V(P') \subseteq W'_{i-1}$ and $V(P') \cap \yset_{i-1} \neq \emptyset$. But then $u,v$ are $\yset_{i-1}$-connected in $G[W'_{i-1}]$, contradiction.

We conclude from the above that $\P$ is a progressive partition of $\mmodel'$.
\qed
\end{proof}

The following lemma addresses the case when $\mmodel$ contains a reducible edge.

\begin{lemma} \label{lem13} 
Suppose that $e \in F$ is a reducible edge in $\mmodel$. Let $\mmodel' = \mmodel / e$. Then $\ppart$ is a progressive partition of $\mmodel'$.
\end{lemma}

\begin{proof} Suppose that $e = xx'$ with $x' = p(x)$ and suppose that $\mmodel' = (X,G',\T')$. Let $Z = \{ v \in V : V(T_v) \cap e \neq \emptyset \}$. We then have $E(G') = \{ uv : uv \in E \text{ or } u,v \in Z \}$. We show that $\ppart$ is a progressive partition of $\mmodel'$. Consider $i \in [p]$. We then find $u,v \in \xset$ that are not $\yset_{i-1}$-connected in $G[W'_{i-1}]$ and are $\yset_i$-connected in $G[W'_{i-1}]$.

First point: $u,v$ are $\yset_i$-connected in $G'[W'_i]$. Since $u,v$ are $\yset_i$-connected in $G[W'_i]$, we find $P$ chordless $(u,v)$-path in $G$ such that $V(P) \subseteq W'_i$ and $V(P) \cap \yset_i \neq \emptyset$. Suppose that $P = z_1 z_2 \ldots z_L$ with $z_1,z_L \in \xset$. If $P$ is chordless in $G'$, we have the desired conclusion. Suppose now that $P$ has a chord in $G'$. We may assume that the chord is $z_p z_q$ with $q-p$ maximum. Hence we have $P' = z_1 z_2 \ldots z_p z_q \ldots z_L$ chordless path in $G'$. By symmetry, we may assume that $V(T_{z_p}) \cap e = \{x\}$ and $V(T_{z_q}) \cap e = \{x'\}$. Since $e$ is reducible, we have $(x,x') \in A_{\mmodel}$ and thus $D^{\T}_x \subseteq D^{\T}_{x'}$. If we had $z_p \in \xset$, we would have $x' \in V(T_{z_p})$, contradiction. Therefore, we have $z_p \in \yset_i$. It follows that $P'$ is a chordless $(u,v)$-path in $G'$ such that $V(P') \subseteq W'_i$ and $V(P') \cap \yset_i \neq \emptyset$. We conclude that $u,v$ are $\yset_i$-connected in $G'[W'_i]$.

Second point: $u,v$ are not $\yset_{i-1}$-connected in $G'[W'_{i-1}]$. By way of contradiction, suppose that we have $P$ chordless $(u,v)$-path in $G'$ such that $V(P) \subseteq W'_{i-1}$ and $V(P) \cap \yset_{i-1} \neq \emptyset$. Suppose that $P = z_1 z_2 \ldots z_L$. If $P$ is a path in $G$, we find that $u,v$ are $\yset_{i-1}$-connected in $G[W'_{i-1}]$, contradiction. Thus, we may assume that $P$ is not a path in $G$. We then find some $p$ ($1 \leq p < L$) such that $z_p, z_{p+1} \in Z$ and non adjacent in $G$. For each $j$ ($j \neq p,p+1$), we have $z_j \notin Z$, as otherwise $z_j$ would be adjacent to $z_p$ and $z_{p+1}$ in $G'$. By symmetry, we may assume that $V(T_{z_p}) \cap e = \{x\}, V(T_{z_{p+1}}) \cap e = \{x'\}$ and $V(T_{z_j}) \cap e = \emptyset$ for each $j \neq p,p+1$. Since $e$ is reducible, we have $D^{\T}_x \subseteq D^{\T}_{x'}$, and thus $z_p \in \yset_{i-1}$. By Lemma \ref{lem9}, we find an edge $e' = xy$ in $T$ with $y \notin e$ such that $S^{\T}_{xy} = (A,S,B)$, $z_p \in S$ and $z_j \in B$ ($1 \leq j < p$). Since $y \notin e$, we have $y \neq x'$ and thus $y$ child of $x$ in $T$. Since $e$ is reducible, we have $D^{\T}_{x} \nsubseteq D^{\T}_{y}$. We then find a vertex $w \in D^{\T}_{x} - D^{\T}_{y}$. Since $D^{\T}_{x} \subseteq D^{\T}_{x'}$, we have $w \in D^{\T}_{x'}$. It follows that $w$ is adjacent in $G$ to $z_p,z_{p+1}$ and nonadjacent in $G$ to $z_j$ ($1 \leq j < p$). Consider $q \geq p+1$ maximum such that $w$ is adjacent in $G$ to $z_q$. Then $P' = z_1 z_2 \ldots z_p w z_q \ldots z_L$ is a chordless $(u,v)$-path in $G$ such that $V(P') \subseteq W'_{i-1}$ and $V(P') \cap \yset_{i-1} \neq \emptyset$. It follows that $u,v$ are $\yset_{i-1}$-connected in $G[W'_{i-1}]$, contradiction.

We conclude from the above that $\P$ is a progressive partition of $\mmodel'$.
\qed
\end{proof}

Equipped with the previous lemmas, we are now in a position to prove Theorem \ref{thm4}.

\begin{theorem} (Theorem \ref{thm4} restated) Let $\mmodel = (X,G,\T)$ be a model and let $\ppart$ be a progressive partition of $\mmodel$. Then $|\ppart| \leq 2|\xset|$.
\end{theorem}

\begin{proof} We may assume that: for every proper minor $\mmodel'$ of $\mmodel$, the ordered partition $\ppart$ is not a progressive partition of $\mmodel'$. As above, we suppose that $\mmodel = (X,G,\T)$, $G = (V,E)$, $\T = (T,\{T_v : v \in V\})$ and $T = (W,F)$. We let $Y = V-X$ and we suppose that $\ppart = (C_1,\ldots,C_p)$ is an ordered partition of $Y$. Suppose that $D_{\mmodel} = (W,A_{\mmodel})$. By Lemmas \ref{lem12} and \ref{lem13}, we have that $D_{\mmodel}$ is acyclic and $\M$ contains no reducible edge.

The following claim provides a linear bound on the number of vertices in the clique tree.

\textit{Claim 1.} $|W| \leq 2|X|+1$.

\textit{Proof.} We let $W_1$ be the set of vertices $x \in W$ with $x \neq r_T$ and $(x,p(x)) \in A_{\mmodel}$, and we let $W_2 = W - W_1$. 

Let us define $\phi : X \rightarrow W$ such that: for each $u \in X$, $\phi(u) = r(u)$. Observe that for each $u \in X$, if $\phi(u) = x$ then $x \in W_2$: indeed, if $p(x) = y \neq \bot$, since $u \in D^{\T}_x - D^{\T}_y$ we have $D^{\T}_x \nsubseteq D^{\T}_y$ and thus $(x,y) \notin A_{\mmodel}$. Moreover, each element $x \in W_2 - r_T$ has a preimage under $\phi$: if $p(x) = y$, since $D^{\T}_x \nsubseteq D^{\T}_y$ we find $u \in D^{\T}_x - D^{\T}_y$ and thus $r(u) = x$. It follows that $|W_2| \leq |X|+1$.

Let us define $\phi' : W_1 \rightarrow W_2$ as follows. Consider $x \in W_1$ with $y = p(x)$, we then have $(x,y) \in A_{\mmodel}$. Since $D_{\mmodel}$ contains no reducible edge, there exists $z$ child of $x$ in $T$ with $(x,z) \in A_{\mmodel}$. Since $D_{\mmodel}$ is acyclic, we have $(z,x) \notin A_{\mmodel}$, and thus $z \in W_2$. Let $\phi'(x) = z$. Observe that $\phi'$ is injective and the image of $\phi'$ is included in $W_2 - r_T$. Thus, we have $|W_1| \leq |W_2|-1$. 

We conclude that: $|W| = |W_1| + |W_2| \leq 2|W_2| - 1 \leq 2|X| + 1$.
\qed

For each $i \in [p]$, let $T_i = \cup_{v \in C_i} T_v$, and let $T'_i = \cup_{j < i} T_j$. The following claim shows that the set $T'_i$ must grow at each step.

\textit{Claim 2.} For each $i \in [p]$, we have $E(T_i) \nsubseteq E(T'_i)$.

\textit{Proof.} By way of contradiction, suppose that $E(T_i) \subseteq E(T'_i)$. Since $\ppart$ is a progressive partition of $H$, we find $u,v \in X$ such that $u,v$ are $\yset_i$-connected in $G[W'_i]$ and not $\yset_{i-1}$-connected in $G[W'_{i-1}]$. Let $\S$ be the set of chordless $(u,v)$-paths $P$ in $G$ such that $V(P) \subseteq W'_i$ and $V(P) \cap \yset_i \neq \emptyset$; we thus have $\S$ nonempty. Let us choose $P \in \S$ such that $|V(P) \cap C_i|$ is minimum. If $V(P) \cap C_i = \emptyset$, we obtain that $u,v$ are $\yset_{i-1}$-connected in $G[W'_{i-1}]$, contradiction. Suppose that $V(P) \cap C_i \neq \emptyset$, and let us choose $w$ in this intersection. Since $u,v \in \xset$, we have $w \in I(P)$. By hypothesis, we have $E(T_w) \subseteq E(T'_i)$. Let $S = \{ z \in Y_{i-1} : E(T_z) \cap E(T_w) \neq \emptyset \}$. By Lemma \ref{lem11}, $G[S]$ is connected and $N_{G}[w] \subseteq N_{G}[S]$. By Lemma \ref{lem10}, we find $P'$ chordless $(u,v)$-path in $G$ such that $V(P') \subseteq V(P) - w + S$ and $I(P') \cap S \neq \emptyset$. We then have $P' \in \S$ and $|V(P') \cap C_i| < |V(P) \cap C_i|$, contradicting the choice of $P$. \qed

Let us complete the proof of the theorem. An immediate induction based on Claim 2 shows that for each $i \in [p]$, $T_i$ has at least $i$ edges. Thus, we have $p \leq |F|$. Since $T$ is a tree, we have $|F| \leq |W|-1$. By Claim 1, we conclude that $p \leq 2|\xset|$.
\qed
\end{proof}

\bibliography{Submission} 
\bibliographystyle{ieeetr}
 
\end{document}